\input epsf
\input harvmac.tex
\def\capt#1{\narrower{
\baselineskip=14pt plus 1pt minus 1pt #1}}

\lref\Muss{Mussardo, G.: Off-critical statistical models: factorizable
scattering theories and bootstrap program. Phys. Rep. {\bf 218}, 215-379
(1992) }

\lref\Wils{Wilson, K.: Non-Lagrangian models of current algebra.
Phys. Rew. {\bf 179}, 1499-1512 (1969)}

\lref\DotsFat{ Dotsenko, Vl.S. and Fateev, V.A.: Conformal algebra and
multipoint correlation functions in 2d statistical models. Nucl. Phys.
{\bf B240}\ [FS{\bf 12}], 312-348 (1984)\semi
Dotsenko, Vl.S. and  Fateev, V.A.: Four-point
correlation functions and the operator algebra in 2d conformal invariant
theories with central charge $c\le1$.
Nucl. Phys. {\bf B251}\ [FS{\bf 13}] 691-734 (1985) }

\lref\luth{Luther, A.: Eigenvalue
spectrum of interacting massive fermions in one dimension.
Phys. Rev. {\bf B14}, 5, 2153-2159 (1976)}

\lref\smbb{Babelon, O., Bernard, D. and Smirnov, F.A.:
Null vectors in integrable field theory.
Commun. Math. Phys. {\bf 182}, 319-354 (1996)}

\lref\Baxter{Baxter, R.J.: Exactly Solved Models
in Statistical Mechanics. London: Academic Press 1982}

\lref\Jons{Johnson, J.D., Krinsky, S. and McCoy, B.M.:
Vertical-arrow correlation length in
the eight-vertex model and the low-lying excitation
of the X-Y-Z Hamiltonian.
Phys. Rev. {\bf A8}, 2526-2547 (1973)}

\lref\ABZ{Zamolodchikov, A.B.:
Integrable field theory from conformal field theory.
Adv. Stud. in Pure Math.
{\bf 19}, 641-674 (1989)}

\lref\Fedya{Smirnov, F.A.: Form-factors in completely
integrable models of
quantum field theory. Singapore: World Scientific 1992}

\lref\lik{Lukyanov S.: Form-factors of exponential fields
in the sine-Gordon model.
Mod. Phys. Lett. {\bf A12}, 2543-2550 (1997) }

\lref\dotse{Dotsenko, V., Picco M. and Pujoi, P.:
Renormalization group calculation of correlation
functions for the 2D random bound Ising and Potts models.
Nucl. Phys. {\bf B455}, 701-723 (1995) }

\lref\gr{Guida, R. and Magnoli, N.: Tricritical Ising model near
criticality. Int. J. Mod. Phys. {\bf A13}, 1145-1158 (1998) }

\lref\ZZ{Zamolodchikov, A.B. and Zamolodchikov, Al.B.:
Structure constants and conformal bootstrap in
Liouville field theory.
Nucl. Phys. {\bf B477}, 577-605   (1996) }

\lref\ZAL{Zamolodchikov, Al.B.: Painlev${\acute {\rm e}}$ and
2D polymers. Nucl. Phys. {\bf B432}, 427-456 (1994)}

\lref\deV{ Destri, C. and  de Vega, H.:
New exact result in affine Toda
field theories: Free energy and
wave functional renormalization.
Nucl. Phys. {\bf B358}, 251-294 (1991)}

\lref\mag{Guida, R. and Magnoli, N.: All order IR finite expansion
for short distance behavior of massless theories perturbed
by a relevant operator.
Nucl. Phys. {\bf B471}, 361-388 (1996)}

\lref\BPZ{Belavin, A.A., Polyakov, A.M. and
Zamolodchikov, A.B.:
Infinite conformal symmetry in two-dimensional
quantum field theory.
Nucl. Phys. {\bf B241}, 333-380 (1984)}

\lref\Leclair{LeClair, A.: Restricted Sine-Gordon theory
and the minimal
conformal series. Phys. Lett. {\bf B230}, 103-107 (1989)}

\lref\Sm{Smirnov, F.A.: Reductions of quantum Sine-Gordon model
as perturbations of Minimal Models of Conformal Field Theory.
Nucl. Phys. {\bf B337}, 156-180 (1990)}

\lref\shifm{Shifman, M.A., Vainshtein, A.I. and Zakharov, V.I.:
QCD and resonance physics. Theoretical foundation.
Nucl. Phys. {\bf B147}, 385-447 (1979)\semi
QCD and resonance physics. Applications.
Nucl. Phys. {\bf B147}, 448-534 (1979)}

\lref\FLZZZ{Fateev, V., Lukyanov, S., Zamolodchikov, A. and
Zamolodchikov, Al.: Expectation values of local fields
in Bullough-Dodd model and integrable
perturbed conformal field theories.
Nucl. Phys. {\bf B516}, 652-674 (1998)}

\lref\Baxter{Baxter, R.J.: Exactly solved models
in statistical mechanics. London: Academic Press 1982}

\lref\Zarn{Zamolodchikov, Al.B.: Mass scale
in the Sine-Gordon model and its
reductions. Int. J. Mod. Phys. {\bf A10}, 1125-1150  (1995) }

\lref\ZamAl{
Zamolodchikov, Al.B.:
Two-point correlation function
in Scaling Lee-Yang model.
Nucl. Phys. {\bf B348}, 619-641 (1991)}

\lref\LZ{Lukyanov, S. and Zamolodchikov, A.:
Exact expectation values of local fields
in quantum sine-Gordon model.
Nucl. Phys. {\bf B493}, 571-587 (1997)}

\lref\FLZZ{Fateev, V., Lukyanov, S., Zamolodchikov, A. and
Zamolodchikov, Al.: Expectation values of boundary fields
in the boundary sine-Gordon model. Phys. Lett. {\bf B406},
83-88 (1997)}

\lref\magnol{Guida, R. and Magnoli, N.: Vacuum expectation values from
a variational approach. Phys. Lett. {\bf B411},
127-133 (1997)}

\lref\SL{Lukyanov, S.: Low energy effective Hamiltonian
 for   the XXZ spin chain. Preprint  RU-97-103, \#cond-mat/9712314 }

\Title{\vbox{\baselineskip12pt
\hbox{RU-98-29}
\hbox{hep-th/9807236}}}
{\vbox{\centerline{
Expectation values of descendent fields}
\vskip6pt
\centerline{in the sine-Gordon  model} }}

\centerline{Vladimir Fateev$^{1,3}$, Dmitri Fradkin$^{2}$, 
Sergei Lukyanov$^{2,3}$,}
\centerline{
Alexander Zamolodchikov$^{2,3}$
and Alexei Zamolodchikov$^{1}$}

\centerline{}
\centerline{$^1$Laboratoire de Physique
Math\'ematique, Universit\'e de Montpellier II}
\centerline{ Pl. E. Bataillon,  34095 Montpellier, FRANCE}
\centerline{$^2 $Department of Physics and Astronomy,
Rutgers University}
\centerline{ Piscataway,
NJ 08855-0849, USA}
\centerline{and}
\centerline{$^3$L.D. Landau Institute for Theoretical Physics,}
\centerline{Chernogolovka, 142432, RUSSIA}

\centerline{}
\centerline{}
\centerline{}
\centerline{\bf Abstract}
\centerline{}
We obtain exactly the vacuum expectation values $\langle\,
(\partial\varphi)^2 ({\bar\partial}\varphi)\, e^{i\alpha\varphi}\, \rangle$
in the sine-Gordon model and $\langle\, 
L_{-2}{\bar L}_{-2} \Phi_{l,k}\, \rangle$ in
$\Phi_{1,3}$ perturbed minimal CFT. We discuss applications of
these results to short-distance expansions of two-point correlation
functions.

\Date{May, 98}

\vfill
\eject

\newsec{Introduction}

\vskip 0.2in

One-point Vacuum Expectation Values (VEV) of local fields are important
characteristics of Quantum Field Theory (QFT) vacuum.
Operator Product
Expansions (OPE)\ \Wils\ give rise to
short-distance expansions for
multipoint correlation functions which involve
the one-point VEV as the basic ingredients\ \refs{\shifm,\ZamAl}.
At the same time the
one-point VEV are nonperturbative objects and no systematic techniques
for their evaluation is known. Some results for these quantities are
available from numerical analyses (see e.g.\ \magnol\ for
numerical results in 2D QFT).
Recently some progress has been made in evaluation of the one-point
VEV in 2D {\it integrable} QFT\ \refs{\LZ,\FLZZ,\FLZZZ}.
In these papers the one-point VEV of the
{\it primary} fields in some integrable QFT, including the sine-Gordon
model and $\Phi_{1,3}$ perturbed minimal CFT, were found exactly. On the
other hand complete characterization of the correlation functions
requires the knowledge of the VEV of all local fields, including the
descendant operators. In the present paper we address the problem of
calculating the one-point VEV of descendant fields.
It was shown in\ \refs{\FLZZ, \FLZZZ}\ that
the VEV of the primary fields in sine-Gordon model (and in similar
integrable QFT) satisfy remarkable ``reflection relation'' which
involves the ``reflection S-matrix'' of Liouville CFT\ \ZZ, and their
one-point VEV can be obtained as appropriate solutions to these
relations. We will show here how this approach can be extended to the
descendant fields and explicitly evaluate the
VEV of the simplest nontrivial
descendants in the sine-Gordon model and in $\Phi_{1,3}$ perturbed
minimal models.

Let us introduce basic notations and state main results of this work.
The sine-Gordon model is defined by the Euclidean action
\eqn\sg{
{\cal A}_{SG} =
\int\ d^2 x\ \biggl\{\
{1\over{16\pi}}(\partial_\nu \varphi)^2
-2\mu\, \cos( \beta\varphi ) \ \biggr\}\ ,}
where $\mu$ and $\beta$ are parameters, $0<\beta^2 < 1$. The simplest
local fields in this QFT are the exponentials $e^{i\alpha\varphi}$.
Exact VEV $\langle\, e^{i\alpha\varphi}\,
\rangle_{SG}$ of these fields are
found in\ \LZ. Here
we will consider more general local fields of the form
\eqn\compos{
(\partial^{n_1}\varphi)\, (\partial^{n_2}\varphi)\,  \ldots\,
(\partial^{n_N}\varphi)\
({\bar\partial}^{m_1}\varphi)\, ({\bar\partial}^{m_2}\varphi)\, \ldots\,
({\bar\partial}^{m_K}\, \varphi)\ e^{i\alpha\varphi}\, ,
\ \ \ \ \ \big|\, \Re e\, \alpha\, \big|<{1\over 2\beta}\, ,  }
where $\partial = \partial_{z},\, {\bar\partial}=\partial_{\bar z}$ and
$z, {\bar z}$ are complex coordinates, $z=x_1+ix_2,\, {\bar z}= x_1 -
ix_2$.
Precise definition of these 
fields in\ \sg\ requires specification of their
renormalizations. We adopt the scheme in which the renormalized
fields\ \compos\ have definite scale dimensions (see e.g.\ \refs{
\ABZ,\ZamAl }). Some details are
presented in Sect.2. In Sect.3 we generalize the ``reflection relations''
of\ \refs{\FLZZ, \FLZZZ}\ to 
the fields\ \compos\ and use these relations to
obtain the one-point VEV of the simplest nontrivial field of this kind,
\eqn\main{\eqalign{
\langle \,(\partial\varphi)^2
({\bar\partial}\varphi)^2& e^{i\alpha\varphi}\,\rangle_{SG} =
-\langle \, e^{i\alpha\varphi}\,\rangle_{SG}\ 
\bigg[\, M\ {\sqrt{\pi}\, 
\Gamma\big({3\over 2}+{\xi\over 2}\big)\over \Gamma\big({\xi\over 2}\big)}
\bigg]^{4}\times\cr
&{\Gamma\big(-{\alpha\xi\over\beta} +{\xi\over 2}\big)\,
\Gamma\big({\alpha\xi\over\beta}+{\xi\over 2}\big)\,
\Gamma\big(-{1\over 2}-{\alpha\xi\over\beta}-{\xi\over 2}\big)\,
\Gamma\big(-{1\over 2}+{\alpha\xi\over\beta}-{\xi\over 2}\big)
\over
\Gamma\big(1+{\alpha\xi\over\beta}
-{\xi\over 2}\big)\,
\Gamma\big(1-{\alpha\xi\over\beta} -{\xi\over 2}\big)\, 
\Gamma\big({3\over 2}+{\alpha\xi\over\beta}
+{\xi\over 2}\big)\,
\Gamma\big({3\over 2}-{\alpha\xi\over\beta} +{\xi\over 2}\big)}\ ,}}
where
\eqn\xibeta{ \xi = {\beta^2\over{1-\beta^2} } }
and $M$ is the sine-Gordon soliton mass which relates to the parameter
$\mu$ in\ \sg\ as\ \Zarn
\eqn\hsydt{
\mu=
{\Gamma(\beta^2)\over\pi\,  \Gamma(1-\beta^2)}\bigg[\, M\, 
{\sqrt{\pi}\,
\Gamma\big({1\over 2}+
{\xi\over 2}\big)\over 2\, \Gamma\big({\xi\over 2}\big)}\,
\bigg]^{2-2\beta^2}\ .}
Let us quote here the simpler form the expression \main \ assumes in
the case $\alpha=0$ \foot{It is interesting to note that Eq.(1.6)
implies remarkably simple relation between the VEV of different fields
associated with the sine-Gordon energy-momentum tensor $T_{\mu \nu}$.
Denoting as usual $T = T_{z z},\  {\bar T} = T_{{\bar z} {\bar z}}$ and
$\Theta = T_{z {\bar z}} = {1\over 4} T_{\nu}^{\nu}$ the irreducible
spin components of $T_{\mu \nu}$ and using the known result for $\langle\,
\Theta\, \rangle_{SG}$\ \deV\ we have
$$
\langle\, T{\bar T} \,\rangle_{SG}=-\langle\, \Theta\,\rangle^2_{SG}\ .
$$},
\eqn\maino{\langle \,(\partial\varphi)^2
({\bar\partial}\varphi)^2\,\rangle_{SG}=
-\pi^2\, M^4 \ 
\tan^2(\pi\xi/2)\ .}

The sine-Gordon QFT is closely related to the Minimal
CFT\ \BPZ\
perturbed by the operator $\Phi_{1,3}$, i.e. the QFT defined by the
action
\eqn\mpaa{{\cal M}_{p/p'}+\lambda\int d^2 x\ 
\Phi_{1,3}(x)\ ,}
where ${\cal M}_{p/p'}$ stands for the formal action of the 
minimal model. Here we consider the massive case $\lambda > 0$. 
As is well known, this CFT can be obtained from\ \sg\ with 
$$\xi ={p\over p'-p}$$
by quantum group restriction\ \refs{\Leclair,\Sm}.
This relation was used in\ \refs{\LZ, \FLZZZ}\ to
obtain the VEV of all primary fields $\Phi_{l,k}$ in the QFT\ \mpaa. 
In a similar fashion, the fields\ \compos\ in\ \sg\ are related
to the descendant fields in\ \mpaa. In particular, the result\ \main\ is
sufficient to derive the expectation values  
of the descendant fields $L_{-2}{\bar
L}_{-2} \Phi_{l,k}$,  namely 
\eqn\evllk{{\langle\, 0_s \mid  L_{-2}{\bar
L}_{-2}  \Phi_{l,k} \mid 0_s \rangle\over
\langle\, 0_s \mid  \Phi_{l,k} \mid 0_s \rangle} =-
\biggl[\, M\ {\sqrt{\pi}\, \Gamma\big(
{3\over  2}+{\xi\over  2}\big)\over 
\Gamma\big({\xi\over  2}\big)}
\, \biggr]^4\  {\cal W}\big(\, (\xi+1) l-
\xi k\,\big)\ ,}
$${\cal W}(\eta)= \xi^{-2}\, (1+\xi)^{-2}\
{\Gamma\big({1+\eta+\xi\over 2}\big)\,
\Gamma\big({\eta-\xi\over 2}\big)\,
\Gamma\big({1+\xi-\eta\over 2}\big)\,
\Gamma\big(-{\eta+\xi\over 2}\big)\over
\Gamma\big({1-\eta-\xi\over 2}\big)\,
\Gamma\big(1+{\xi-\eta\over 2}\big)\,
\Gamma\big({1-\xi+\eta\over 2}\big)\,
\Gamma\big(1+{\eta+\xi\over 2}\big)}\ .$$
Here $\mid 0_s \rangle,\  s=1, 2, \ldots , p-1$ is any one of $p-1$
degenerate ground states of the QFT \mpaa, and $M$ is the mass of its
fundamental kink. This mass is related to $\lambda$ in \mpaa\  as \ \Zarn
\eqn\ksjdtr{\lambda^2  ={ (1+\xi)^4\over (1-\xi)^2\, (1-2\xi)^2\, \pi^2}\ 
{\Gamma\big({\xi\over 1+\xi}\big)\, 
\Gamma\big({3\xi\over 1+\xi}\big)\ \over
\Gamma\big({1\over 1+\xi}\big)\, \Gamma\big({1-2\xi\over 1+\xi}\big)}\
\bigg[\, M\,
{\sqrt{\pi}\,
\Gamma\big({1\over 2}+
{\xi\over 2}\big)\over 2\, \Gamma\big({\xi\over 2}\big)}\,
\bigg]^{{4\over 1+\xi}}\ .}

In fact, the ``reflection relations'' admit certain ambiguity to their
solution, akin to the ``CDD ambiguity'' in the factorizable S-matrix
theory (see e.g. \Muss ). To some extent the ambiguity can be narrowed by
taking into account the ``resonance conditions'' (see Sect.2). In
\main\ we have fixed this ambiguity by choosing the ``minimal 
solution'' - the simplest solution compatible with the resonance
conditions (choosing the minimal solution is a common practice in the
S-matrix theory). This choice is confirmed in Sect.4-5, where \main\
is checked against results obtained in \sg\ using
semiclassical approximation
(Sect.4) and ordinary Feynmann perturbation theory (Sect.5). Moreover,
the special case \maino\ can be obtained directly from exact lattice
theory of the XYZ model, as we show in Sect.6. Finally, in Sect.7 we use
the Eq.\evllk\ (more precisely, its particular case $l=k=0$) to extend
by one more order the short-distance expansion of the two-point
correlation function of the Scaling Lee-Yang Model \ZamAl.

\newsec{ Descendant fields and Operator Product Expansions}

\vskip 0.2in

The sine-Gordon model\ \sg\ can be regarded as Gaussian CFT
\eqn\gauss{{\cal A}_{Gauss}=
\int d^2 x \ {1\over {16\pi}}\, (\partial_{\nu}
\varphi)^2}
perturbed by the relevant operator $2\cos(\beta\varphi)$. Let ${\cal
F}_{Gauss}$ be the space of local fields in the CFT\ \gauss. This space
is spanned by the fields\ \compos. In the free field theory\ \gauss\ these
composite fields are defined through usual Wick ordering with
$$\langle\,  \varphi(z, {\bar z})\, \varphi (0, 0)\, \rangle_{Gauss} =
-2\, \log(z{\bar z})\ .$$
With this definition the field\ \compos\ has the
dimensions $(\Delta, {\bar\Delta})=(\alpha^2 + l, \alpha^2+{\bar
l}\,)$, where the integer ``levels''\ $l,\,  {\bar l}$\ are the
sums  $l= \sum_{i=1}^{N}
n_i,\  {\bar l} = \sum_{j=1}^K m_j$.
The sum
\eqn\dimmm{D= \Delta+{\bar\Delta} = 2\alpha^2 + l+{\bar l} }
is
the scale dimension of the field\ \compos\ while
the difference $S=l-{\bar l}$
coincides with its spin. Note that some linear combinations of the
fields\ \compos\ are total
derivatives of other local fields, for example
\eqn\deriva{
-\alpha^2\,  (\partial\varphi)({\bar\partial}\varphi)\, e^{i\alpha\varphi}=
\partial{\bar\partial}e^{i\alpha\varphi}\ ,}
\eqn\derivb{\eqalign{
&(\partial^2\varphi)({\bar\partial}\varphi)^2 e^{i\alpha\varphi} +i\alpha
(\partial\varphi)^2({\bar\partial}\varphi)^2 e^{i\alpha\varphi}=
\partial\big((\partial\varphi)({\bar\partial}\varphi)^2
e^{i\alpha\varphi}\big)\, ,\cr  
&(\partial^2\varphi)({\bar\partial}^2\varphi)\, e^{i\alpha\varphi} +i\alpha
(\partial\varphi)^2({\bar\partial}^2\varphi)\, e^{i\alpha\varphi}=
\partial\big((\partial\varphi)({\bar\partial}^2\varphi)\,
e^{i\alpha\varphi}\big)\ .}}
These elementary relations follow from
the equation of motion of\ \gauss,
\eqn\ff{\partial{\bar \partial}\varphi =0\ .}

The perturbation in\ \sg\ leads to additional divergences in the matrix
elements of
the fields\ \compos, and so certain counterterms have to be added
to \compos\ in order to compensate 
for these divergences. These counterterms
contain local fields of the same spin, with cutoff-dependent
coefficients. However, as long as the perturbation is relevant
(i.e. $0<\beta^2 < 1$), the situation is relatively simple. For a given
field\ \compos\ it suffices to add only the fields of lower scale 
dimensions. In particular for a given
field\ \compos\ there are finitely many counterterms. As usual these
counterterms are not completely determined 
by the sole requirement that they
absorb all the divergences, there is always a possibility to add finite
counterterms. However, except for when certain resonance conditions are
satisfied (see below), the ambiguity is eliminated completely if one
demands that the resulting renormalized field (i.e. the unperturbed
field\ \compos\ plus the counterterms) has definite scale dimension,
which then coincides with\ \dimmm. We say that the field ${\cal O}_i$
has $n$-th order resonance with the field ${\cal O}_j$ if the dimensions
of these fields satisfy the equation
\eqn\reswa{ D_i = D_j + 2 n\, (1-\beta^2)\ }
with some positive integer $n$.
If this resonance condition is satisfied an obvious ambiguity
\eqn\amb{{\cal O}_i \to {\cal O}_i + {\rm const}\  \mu^n\, {\cal O}_j }
in defining the renormalized field ${\cal O}_i$ with the scale
dimension $D_i$ typically results in the logarithmic
scaling of ${\cal O}_i$.

We define the field\ \compos\ in the perturbed theory\ \sg\ as
the renormalized field satisfying the following conditions:
i) \compos\  has definite scale dimension\ \dimmm\ and ii) the
short-distance limit of its correlations coincides with the correlations
of the corresponding field\ \compos\ in CFT\ \gauss.
As explained above,
in non-resonant cases this definition is unambiguous. In resonant cases
it is not complete and one has to impose additional conditions to fix
the ambiguity\ \amb. For a given field\ \compos\ the 
resonances\ \reswa\ occur
at isolated values of $\alpha$. In this paper we are
interested in generic values of this parameter and so we will not
elaborate any specific convention concerning resonant cases. It
suffices to note that if the matrix elements of\ \compos\ are viewed
as the functions of $\alpha$ the resonances show up
as the poles in this variable.

It is important to note that under this definition of
the fields\ \compos\ in\ \sg\ these fields satisfy exactly
the same relations\ \deriva,\ \derivb\ (as
well as similar higher-level relations) as the fields in\ \gauss. In
this sense the symbol\ $\varphi$\ in\ \compos\ is the subject to the
``free'' equation of motion\ \ff\ rather then to the full equation of
motion of the sine-Gordon theory. This is not a contradiction because 
it is rather the unrenormalized fields in\ \sg\ that satisfy the full 
equation of motion; the renormalized ones differ from those by
counterterms. In this paper we are interested in the one-point VEV of
the fields\ \compos\ in the QFT\ \sg. Let us mention here some elementary
properties of these VEV. First, only the fields of zero spin, i.e. with
$l={\bar l}$, have nonvanishing one-point VEV. Next, the one-point VEV
of the fields which are total derivatives vanish.
According to\ \deriva,\ \derivb\ the
only spinless field of the level $l=1$ has vanishing VEV,
\eqn\totala{\langle\,
(\partial\varphi)({\bar\partial}\varphi)\, e^{i\alpha\varphi}\,
\rangle_{SG}=0\, , }
There are four linearly independent spinless fields on the level $l=2$.
However the following relations among their VEV are simple consequences
of\ \derivb,
\eqn\totalb{\eqalign{&\langle\, (\partial^2\varphi)\,
({\bar\partial}^2\varphi)
e^{i\alpha\varphi}\, \rangle_{SG} = -i\alpha\ \langle\,
(\partial\varphi)^2({\bar\partial}^2\varphi)
\, e^{i\alpha\varphi}\, \rangle_{SG}=\cr
&-i\alpha\ \langle\,
(\partial^2\varphi)({\bar\partial}\varphi)^2\, e^{i\alpha\varphi}
\, \rangle_{SG} = -\alpha^2\ \langle
\, (\partial\varphi)^2({\bar\partial}\varphi)^2\, e^{i\alpha\varphi}\,
\rangle_{SG}\ .}}
These relation allow one to express all $l=2$ VEV through the VEV
\eqn\vvv{
\langle\, (\partial\varphi)^2({\bar\partial}\varphi)^2 e^{i\alpha\varphi}\,
\rangle_{SG}\ .}
Starting from the level $l=3$ there are additional
``kinematic'' relation among the VEV following from the existence of
higher local Integrals of Motion in the QFT\ \sg\ \smbb, but we will not
discuss them here. Instead we will concentrate attention on 
the VEV\ \vvv.

Let us make here a simple remark 
concerning the properties of the VEV\ \vvv\ as the function of $\alpha$.
It is easy to check that the field 
$(\partial\varphi)^2({\bar\partial}\varphi)^2 e^{i\alpha\varphi}$ has 
a second order resonance with the field $e^{i(\alpha+2\beta)\varphi}$
at $\alpha=-\beta/2$. Similarly, at\ $\alpha = \beta/2$\ it has second
order resonance with $e^{i(\alpha-2\beta)\varphi}$.
Therefore the VEV\ \vvv\ is expected to have 
poles at $\alpha = \pm\beta/2$.
As we will argue below the
residues at these poles can be expressed through the VEV of the primary
fields responsible for the resonances.

To explain this point let us consider a product of two primary 
fields $e^{i\alpha_1 \varphi}(x)\,e^{i\alpha_2 \varphi}(y)$ in\ \sg.
The corresponding OPE has the form
\eqn\ope{e^{i\alpha_1 \varphi}(x)\,e^{i\alpha_2 \varphi}(y) =
\sum_{n=-\infty}^{\infty} \bigg\{\, C_{\alpha_1 \alpha_2}^{n,0}
( r)\ e^{i(\alpha + n\beta)\varphi}(y) +
\ldots\, 
\bigg\}\ ,}
where $\alpha=\alpha_1 + \alpha_2$, $r=|x-y|$,\ and the dots in each term
stand for contributions of the descendants\ \compos\ of the field 
$e^{i(\alpha + n\beta)\varphi}(y)$.
The coefficient functions $C$ are in principle computable within the 
Conformal Perturbation Theory (CPT)\ \ZamAl\ (see also\ \mag).
The CPT suggests
for them the following form 
\eqn\cform{C_{\alpha_1 \alpha_2}^{n,0}( r)=\mu^{|n|}\ 
r^{4\alpha_1 \alpha_2+4 n\beta (\alpha_1 +\alpha_2)+
2|n|(1-\beta^2)+2 n^2\beta^2}
\ f_{\alpha_1 \alpha_2}^{n,0}\ \big(\mu^2
r^{4-4 \beta^2}\big)\ ,}
where the functions $f$ in\ \cform\ admit power series expansions, i.e.
\eqn\fexp{f_{\alpha_1 \alpha_2}^{n,0}(t)=
\sum_{k=0}^{\infty}\ f_{k}^{n,0}(\alpha_1, \alpha_2)\ t^k\ .}
The CPT gives the coefficients in\ \fexp\ in terms
of certain\ $2\, |n|+2\, k$--fold
Coulomb-type integrals. Note that
the leading terms $f_{0}^{n,0}(\alpha_1, \alpha_2)$ in the
series\ \fexp\ are expressed through the integrals
\eqn\fatey{j_n(a, b,\rho)=
{1\over n!}\ \int \prod_{k=1}^n\ d^2x_k\
\prod_{k=1}^n\ |x_k|^{4a}\, |1-x_k|^{4b}\, 
\prod_{k<p}^n\ |x_k-x_p|^{4\rho}\ ,}
namely
\eqn\jshdyy{\eqalign{&f_{0}^{0,0}(\alpha_1, \alpha_2) =1\, ,\cr
&f_{0}^{n,0}(\alpha_1, \alpha_2) = j_n \big(\alpha_1\beta,
\alpha_2\beta,\beta^2 \big) \quad {\rm for}\ \  n>0\, ,\cr
&f_{0}^{n,0}(\alpha_1, \alpha_2) =
j_n \big(-\alpha_1\beta,
-\alpha_2\beta,\beta^2 \big)  \quad {\rm  for}\ \  n<0\, .}}
The integrals\ \fatey\ are evaluated explicitly\ \DotsFat,
\eqn\fateyd{\eqalign{j_n(a, b,&\rho)=
\pi^n \ \big(\gamma(\rho)\big)^{-n}\ 
\prod_{k=1}^n\
\gamma(k\rho)\times\cr
&\prod_{k=0}^{n-1} \gamma(1+2a+k\rho)\, \gamma(1+2b+k\rho)\,
\gamma\big(-1-2a-2b-(n-1+k)\rho\big)\ .}}
Here and below the notation \ $\gamma(t)={\Gamma(t)/\Gamma(1-t)}$ is used.
Let us quote also the expression for the first subleading term in the
expansion\ \cform\ of the function $C_{\alpha_1 \alpha_2}^{0,0}$,
\eqn\gsdfed{
f_{1}^{0,0}(\alpha_1, \alpha_2)=J(\alpha_1 \beta, \alpha_2 \beta ,
\beta^2)\ ,}
where 
\eqn\Jint{
J(a,b,\rho)=
\int d^2 x\, d^2 y\ |x|^{4 a}\,
|y|^{-4 a}\, |1-x|^{4 b}\, |1-y|^{ -4 b}\,
|x-y|^{-4\rho}\ .
}
This integral can be expressed through generalized 
hypergeometric function\ ${}_3F_2$\ at unity\ \dotse\ (see also\ \gr).
The coefficient functions standing in front of the 
descendant field in\ \ope\ admit similar CPT expansions.

There are reasons to believe that the series\ \fexp\ (and similar CPT
series for the coefficient functions corresponding to the descendant
fields in\ \ope) converge for all complex $t$. But independently of
the convergence these series can be used to generate asymptotic 
short-distance expansion for the two-point correlation function
\eqn\twopoint{
{\cal G}_{\alpha_1 \alpha_2}(r) = \langle\, e^{i\alpha_1 \varphi}(x)
e^{i\alpha_2 \varphi}(y)\, \rangle_{SG}\, , \qquad r=|x-y|\ ,}
provided the one-point VEV of the exponential fields in
the r.h.s. of\ \ope, and also the one-point VEV 
of their descendants\ \compos, are
known. For the exponential fields the one-point VEV
\eqn\gaa{{\cal G}_{\alpha}=\langle\, e^{i\alpha\varphi}\, \rangle_{SG}}
closed analytic expression exists\ \LZ.
According to our discussion above
the first nonzero contribution due to the descendants comes from the fields 
$(\partial\varphi)^2 ({\bar\partial}\varphi)^2
e^{i(\alpha+n\beta)\varphi}$, namely
\eqn\opea{{\cal G}_{\alpha_1 \alpha_2}(r) = \sum_{n=-\infty}^{\infty}
\bigg\{ C_{\alpha_1 \alpha_2}^{n,0}(r)\,
\langle\, e^{i(\alpha+n\beta)\varphi}\, \rangle_{SG} +
C_{\alpha_1 \alpha_2}^{n,2}(r)\, 
\langle\, (\partial\varphi)^2 ({\bar\partial}\varphi)^2 
e^{i(\alpha+n\beta)\, \varphi}\,\rangle_{SG}+\ldots \bigg\}\, ,}
where $\alpha=\alpha_1+\alpha_2$ and the omitted 
terms contain the descendants of the levels $l={\bar
l} = 4$ or higher. The coefficient functions $C_{\alpha_1
\alpha_2}^{n,2}(r)$ admit CPT expansions similar to\ \cform,\ \fexp.
In particular,
\eqn\sjdht{C_{\alpha_1 \alpha_2}^{0,2}(r)=
-{{(\alpha_1 \alpha_2)^2}\over 4}\
r^4\  \Big(1 + O\big(\mu^2 r^{4-4\beta^2}\big)\, \Big)\ .}
Combining all these expressions one can write down the $r\to 0$
expansion
\eqn\reewwe{\eqalign{&{\cal G}_{\alpha_1 \alpha_2}(r)=
{\cal G}_{\alpha_1+\alpha_2} \ r^{4 \alpha_1\alpha_2}\
 \Big\{1+
J(\alpha_1\beta, \alpha_2\beta, \beta^2)\  \mu^2\ r^{4-4\beta^2}-
{(\alpha_1\alpha_2)^2\over 4}\ {\cal H}(\alpha_1+\alpha_2)\ r^4\cr
&+
O\big(\mu^4\, r^{8-8\beta^2}\big)\ \Big\}+
\sum_{n=1}^{\infty}\  {\cal G}_{\alpha_1+\alpha_2+
n\beta}\ j_n(\alpha_1\beta, \alpha_2\beta, \beta^2)\ 
\Big\{1+O\big(\mu^2\, r^{4-4\beta^2}\big)\, \Big\}\times\cr
& \mu^n\,  r^{4 \alpha_1\alpha_2 + 4 n\beta (\alpha_1+\alpha_2)
+ 2 n (1-\beta^2)+
2n^2\beta^2}+ 
\sum_{n=1}^{\infty}\ {\cal G}_{\alpha_1+\alpha_2-
n\beta}\ j_n(-\alpha_1\beta,- \alpha_2\beta, \beta^2)\times\cr
& \Big\{1+O\big(\mu^2\, r^{4-4\beta^2}\big)\, \Big\}
\ \mu^n\   r^{4 \alpha_1\alpha_2 - 4 n\beta (\alpha_1+\alpha_2)
+ 2 n (1-\beta^2)+
2n^2\beta^2}\ ,}} 
where ${\cal H}(\alpha)$ stands for the ratio
\eqn\halpha{
{\cal H}(\alpha)={\langle\,
(\partial\varphi)^2({\bar\partial}\varphi)^2 e^{i\alpha\varphi}\,
\rangle_{SG}
\over\langle\,  e^{i\alpha\varphi}\, \rangle_{SG}  }\ .
}
Note that at $\alpha = -\beta/2$ the leading contribution of the field 
$e^{i(\alpha+2\beta)\varphi}$ has the same power low in $r$ as the
contribution of the descendant\ $(\partial\varphi)^2 
({\bar\partial}\varphi)^2
e^{i\alpha\varphi}$. This is exactly the second order
resonance we have mentioned above. The contribution comes with the 
coefficients $j_{2}$ which exhibit the pole at this value of $\alpha$ as
is seen from\ \fateyd. The VEV\ \vvv\ also has
a resonance pole at this point,
and these two pole terms must compensate. This requirement
leads to the relation
\eqn\ress{{\rm res}_{\alpha=-\beta/2}{\cal H}(\alpha)= \bigg(
{\pi\, \mu\over \gamma(\beta^2)} \bigg)^{2+2\xi}\ 
{4\over \beta}\ (1+\xi)^3\, \gamma(-{1\over 2}-\xi)\, \gamma(\xi)\ .}
The compensation of the poles at $\alpha = -\beta/2$ results in the
logarithmic term $r^{4\alpha_1 \alpha_2+4}\log(r)$ in the short distance
expansion of\ \twopoint\ with $\alpha_1 + \alpha_2 = -\beta/2$. The
relation\ \ress\ will 
be used in the next section to fix the normalization of
the VEV\ \vvv.

\newsec{Reflection relations}

\vskip 0.2in

In\ \FLZZZ\ the ``reflection property'' of the
Liouville CFT was used to 
derive the one-point VEV of the exponential fields
$e^{i\alpha\varphi}$\ in sine-Gordon model.
Let us first briefly remind the arguments of
\FLZZZ, and then show how these arguments
can be extended to the case of the
descendant fields\ \compos.

The sine-Gordon model\ \sg\ is closely related to the sinh-Gordon model
\eqn\shg{
{\cal A}_{shG} =
\int\ d^2 x\ \biggl\{\
{1\over{16\pi}}(\partial_\nu \varphi)^2
+2\mu\, \cosh( b\varphi ) \ \biggr\}\ .}
In particular, the one-point VEV $\langle\,
e^{i\alpha\varphi}\, \rangle_{SG}$ and $ \langle\,
e^{a\varphi}\, \rangle_{shG}$
in the two models are related through the substitution
\eqn\none{b=i\beta\, , \qquad a=i\alpha\ .}
This relation also holds for the one-point VEV of the descendant
fields\ \compos. In turn, the sinh-Gordon model can be regarded as 
the Liouville CFT,
\eqn\liouv{
{\cal A}_{L} =
\int\ d^2 x\ \biggl\{\
{1\over{16\pi}}(\partial_\nu \varphi)^2
+\mu\, e^{ b\varphi } \ \biggr\}\ ,}
perturbed by the operator $e^{-b\varphi}$. As is known\ \ZZ, the
correlation functions of the fields $e^{a\varphi}$ in the Liouville
theory exhibit the following ``reflection property'',
\eqn\rprop{\langle\, e^{a\varphi}(x)\, \ldots\, \rangle_{L} = 
R(a)\ \langle\, e^{(Q-a)\varphi}(x)\, \ldots\, \rangle_{L}\ ,}
where
\eqn\qqq{Q=b^{-1}+b}
and the coefficient function
\eqn\ksdu{R(a)=
-\bigg({{\pi\mu\, \Gamma(b^2)}\over{\Gamma(1-b^2)}}\bigg)^{
1+{1\over b^2}- {2a\over b}}\
{{\Gamma\big(-{1\over b^2}+{2a\over b}\big)\,
\Gamma\big(-b^2+2 a b\big)}
\over{\Gamma\big(2+{1\over b^2}-{2a\over b}\big)\,
\Gamma\big(2+b^2-2 a b\big)}}  }
is essentially the vacuum reflection amplitude of the Liouville
CFT. The relation\ \rprop\ suggest the following ``reflection relation''
for the one-point VEV of\ \shg,
\eqn\rrel{
\langle\,  e^{a\varphi}\, \rangle_{shG}=
R(a)\ \langle\,  e^{(Q-a)\varphi}\, \rangle_{shG}\ .}
Combining this relation with the obvious symmetry property
\eqn\symm{
\langle\,  e^{a\varphi}\, \rangle_{shG}=
\langle\,  e^{-a\varphi}\, \rangle_{shG}\ }
and certain assumptions about analytic properties
of\ $\langle\,  e^{a\varphi}\, \rangle_{shG}$\ \FLZZZ\ one
can derive VEV for exponential field.

Conformal symmetry of the Liouville theory\ \liouv\ is generated by the
energy-momentum tensor
\eqn\em{\eqalign{&
T_L(z)=-{1\over 4}\,
(\partial\varphi)^2 +{Q\over 2}\, \partial^2 \varphi\ ,
\cr
&{\bar T}_L({\bar z})=-{1\over 4}\, ({\bar\partial}\varphi)^2 +
{Q\over 2}\, {\bar\partial}^2 \varphi\ .}}
The exponentials $e^{a\varphi}$ are primary fields with respect to the
Virasoro algebra generated by\ \em. Let us introduce the notation
\eqn\virdesc{L_{[n]}\,
{\bar L}_{[m]}\, e^{a\varphi} \equiv
L_{-n_1}L_{-n_2}\, \ldots\, L_{-n_N}{\bar L}_{-m_1}{\bar L}_{-m_2}\,
\ldots\, {\bar
L}_{-m_K}\, e^{a\varphi} }
for the corresponding descendant fields. The symbols
$[n]$ and $[m]$ here stand for arbitrary strings $[-n_1 , -n_2 ,
\ldots ,-n_N]$, $[-m_1,-m_2, \ldots , -m_K]$.
In\ \virdesc\ $L_{n}, {\bar
L}_{n}$ are standard Virasoro generators associated
with\ \em. It is possible to show that the reflection property extends
to all these descendants, namely
\eqn\rpropp{\langle\, 
L_{[n]} {\bar L}_{[m]}\, e^{a\varphi}(x)\, \ldots\, \rangle_{L} = 
R(a)\ \langle\,
L_{[n]} {\bar L}_{[m]}\, e^{(Q-a)\varphi}(x)\, \ldots\, \rangle_{L}\ .}
The arguments identical to those in\ \FLZZZ\ suggest then that the
``reflection relation''\ \rrel\ generalizes to the descendant
fields\ \virdesc\ in straightforward way,
\eqn\rrr{\langle\,  L_{[n]}{\bar L}_{[m]}\, 
e^{a\varphi}\, \rangle_{shG} =R(a)\
\langle\,  L_{[n]}{\bar L}_{[m]}\,
e^{(Q-a)\varphi}\, \rangle_{shG}\ .}
The generalization of the symmetry relation\ \symm\ is less
straightforward. The relation\ \symm\ is a simple consequence of the
symmetry $\varphi\to -\varphi$ of the action\ \shg. However, while the
action\ \shg\ is invariant with respect\ this transformation, the
components\ \em\ of 
the modified energy-momentum tensor, and hence the corresponding
Virasoro generators $L_{n}, {\bar L}_n$, are not. In this respect
the basis
\eqn\composs{
(\partial^{n_1}\varphi)\, (\partial^{n_2}\varphi)\,  \ldots\,
(\partial^{n_N}\varphi)\
({\bar\partial}^{m_1}\varphi)\, ({\bar\partial}^{m_2}\varphi)\, \ldots\,
({\bar\partial}^{m_K}\, \varphi)\ e^{a\varphi}  }
in the space of the descendants is more convenient as
the fields\ \composs\ transform under this reflection in an obvious way.
The fields\ \virdesc\ can be written down as the linear combinations of
the fields\ \composs\ of the same levels\ $l,\, {\bar l}$\ and vice
versa \foot{Of course if the Virasoro module with the primary field
$e^{a\varphi}$ has a null vector at the level $l$ the relation between
\virdesc\ and \composs\ becomes singular. Below we consider generic case
of $a$ and ignore this subtlety.}.
Finding this relation for given levels requires solving a finite
algebraic problem, as explained in\ \ZZ. Here we will only need the
relation\ \ZZ,
\eqn\ttbar{L_{-2}{\bar L}_{-2}\, e^{a\varphi}=
\Big(-{1\over 4}\, (\partial\varphi)^2+
\big({Q\over 2}+a\big)\, \partial^2 \varphi\, \Big)
\Big(-{1\over 4}\, ({\bar\partial}\varphi)^2+
\big({Q\over 2}+a\big)\, {\bar\partial}^2\varphi\, \Big)\, 
e^{a\varphi}\ . }
Consider the one-point VEV of the field\ \ttbar,
\eqn\lphi{\langle\,  L_{-2}{\bar L}_{-2}\, e^{a\varphi}\, \rangle_{shG}=
{1\over 16}\ \big(1+2a\, (Q+2a)\big)^2\  \langle\,
(\partial\varphi)^2({\bar\partial}\varphi)^2\,
e^{a\varphi}\, \rangle_{shG}\ ,}
where the relations analogous to\ \totalb\ were
used to simplify the r.h.s. Then it
follows from\ \rrr\ that
\eqn\jshdyt{\eqalign{
\big(1+2a\, (Q+2a)\big)^2\ &\langle\, (\partial\varphi)^2
({\bar\partial}\varphi)^2\,
e^{a\varphi}\, \rangle_{shG} =\cr
& R(a)\ 
\big(1+2\, (Q-a)(3Q-2a)\big)^2\  \langle\,
(\partial\varphi)^2({\bar\partial}\varphi)^2\,
e^{(Q-a)\varphi}\, \rangle_{shG}\ .}}
We find that the function
\eqn\ha{{ H}(a)=
{\langle\, (\partial\varphi)^2({\bar\partial}\varphi)^2
\, e^{a\varphi}\, \rangle_{shG} \over
\langle\, e^{a\varphi}\, \rangle_{shG} } }
satisfies the functional equations
\eqn\feqa{\eqalign{&H(a) = 
\bigg[\, {(2b+3/b-2a)(3b+2/b-2a)\over (b+2a)(1/b+2a)}\, \bigg]^2\
H(Q-a)\, ,\cr 
&H(a)=H(-a)\ ,}}
where the second equation follows
from the obvious symmetry of\ \ha.
Note that the equation\ \feqa\ remains unchanged if one makes the 
substitution  
\eqn\dual{b\to b^{-1}\ .}
This is in agreement with well known ``duality'' symmetry of the
sinh-Gordon model\ \shg, which in particular implies that all VEV 
of the fields\ \composs\ must be invariant with respect
to the transformation\ \dual.

Obviously, the equations\ \feqa\ determine the function $H(a)$ only up to
a factor $F(a)$ which is an even periodic function,
\eqn\cdd{F(a) = F(-a)\, , \qquad F(a)=F(a+Q)\ .}
The solution we are interested in must have the poles at $a=\pm b/2$
corresponding to the second order resonances discussed in the previous
section. Also, the function $H(a)$ must respect the symmetry\ \dual.
Strictly speaking, this information is not sufficient to fix the
ambiguity\ \cdd. Nevertheless, there is a ``minimal'' solution which
satisfies the above requirements,
\eqn\mainn{H(a)=-\bigg[\, {m \, \Gamma\big({b\over 2 Q}\big)\, 
\Gamma\big({1\over 2 b Q}\big)\over 8 Q^2\ \sqrt\pi}\,  \bigg]^4\,
\gamma\big({a\over Q}-{b\over {2Q}}\big)\, 
\gamma\big(-{a\over Q}-{b\over {2Q}}\big)\, 
\gamma\big({a\over Q}-{1\over
{2bQ}}\big)\,
\gamma\big(-{a\over Q}-{1\over {2bQ}}\big)\, ,}
where $\gamma(t)={\Gamma(t)/\Gamma(1-t)}$ and $m$ is
the mass of the sinh-Gordon particle. 
The residue condition analogous to\ \ress\ is used to fix the overall
normalization of\ \mainn. 
We conjecture that this minimal solution gives exact
ratio\ \ha\ in the sinh-Gordon model. 
The VEV\ \vvv\ is then obtained by the
substitution\ \none, which yields\ \main. In the
subsequent sections we give some evidence in support of this conjecture.

\newsec{Comparison to semiclassical results}

\vskip 0.2in

The result \main\ can be checked against certain semiclassical
calculations in \sg. Consider the two-point correlation function 
\twopoint\ with 
\eqn\alphas{\alpha_1 = \omega\beta\, , \qquad \alpha_2 = \sigma/\beta\ ,} 
where both $\sigma, \omega \sim 1$, in the limit $\beta\to 0$.  In this
limit the functional integral defining \twopoint\ is dominated by the 
saddle-point configuration $\varphi_{cl}(x) = {{2i}\over\beta}\,\phi(t),
\ \ t=m\,|x-y|$,
where $\phi(t)$ is a solution to the Painlev\'e III equation
\eqn\penIII{\partial_t^2\phi+t^{-1}\, \partial_t \phi ={1\over 2}\
\sinh\big(2 \phi\big)\, }
regular at $t>0$ and satisfying the asymptotic conditions 
\eqn\penass{\eqalign{
& \phi(t)=2\sigma\  \log(8/t)-\log\Big(\gamma\big({1\over 2}-\sigma\big)
\Big)+O\big(t^{2\pm 4\sigma}\big)
\ \ \ \ \ \ {\rm as }\ \ t\to 0\, , \cr
&\phi(t)\to  {2 \sin(\pi \sigma)\over \pi}\ K_0(t)
\ \ \ \ \ \ {\rm as}\ \ t\to +\infty\ ,  }}
where $K_0(t)$ is the MacDonald function and again
$\gamma(x)=\Gamma(x)/\Gamma(1-x)$. Therefore the correlation function under
consideration can be written as
\eqn\semcorr{{{\langle\,
e^{i\omega\beta\varphi}(x)\,
e^{i{\sigma\over\beta}\varphi}(y)\,\rangle_{SG}}
\over{\langle\, e^{i\omega\beta\varphi}\,\rangle_{SG}\ \langle\, 
e^{i{\sigma\over\beta}\varphi}\, \rangle_{SG}}}\bigg|_{\beta^2\to 0}=
\big(e^{2\phi(t)}\big)^{-\omega}\ .}
As is known \ZAL, this solution to the Painlev\'e III equation admits
a double-series expansion
\eqn\double{e^{2 \phi(t)}={1\over 4}\ \sum_{m,n=0}^{\infty}\
\big( m+n+2\sigma\, (m-n)\big)^2\ B_{m,n}\ \Big({t\over 8}\Big)^{
2 (m+ n-1+2\sigma(m-n))}\ ,} 
where the coefficients $B_{m,n}$ satisfy certain recursion relations
(see \ZAL\ for details). Using these relations one can derive
explicitly few first terms of the expansion \double,
\eqn\penexp{\eqalign{&e^{2\phi(t)}=
\gamma^2\big({1\over 2}+\sigma\big)\ \Big({t\over 8}\Big)^{-4\sigma}+
{ 8\over (1-2\sigma)^2}\ \gamma^4\big({1\over 2}+\sigma\big)\
\Big({t\over 8}\Big)^{2-8\sigma}-\cr
&\ \ \ \ \ \ \ \ \ \ {8\over
(1+2\sigma)^2}\ \Big({t\over 8}\Big)^{2}+
{48\over (1-2\sigma)^4}\ \gamma^6\big({1\over 2}+\sigma\big)
\ \Big({t\over 8}\Big)^{4-12\sigma}-\cr
&{64\, (1-2\sigma)
\over  (1-4\sigma^2)^2}\ \gamma^2\big({1\over 2}+\sigma\big)\
\Big({t\over 8}\Big)^{4-4\sigma}+
{16\over(1+2\sigma)^4}\ \gamma^2\big({1\over 2}-\sigma\big)\
\Big({t\over 8}\Big)^{4+4\sigma}+\,
 O\big(t^{6-16\sigma}, t^6\big) \ .}}
This expansion is to be compared with the corresponding limiting case
of the expansion \reewwe. To make this comparison straightforward one
can use the the relation
\eqn\vevcl{{{\cal G}_{\sigma/\beta+\omega\beta+n\beta}
\over {\cal G}_{\sigma/\beta}\
{\cal G}_{\omega\beta}}\Big|_{\beta^2\to 0}\to \
\Big({m\over 8}\Big)^{4(\omega+n) \sigma}\ \Big[\gamma\big({1\over 2}-
\sigma\big)\Big]^{2\omega+2 n}\ ,}
which is obtained from the explicit formula for the VEV \gaa\ \LZ , and
the following limiting expressions for the integrals  \Jint\   and
\fateyd\ 
\eqn\intcl{\eqalign{&J(\sigma, \omega\beta^2, \beta^2)\big|_{\beta^2\to 0}
\to -8\pi^2\, \beta^4\  {\omega\, (\sigma+\omega)
\over (1-4\sigma^2)^2}\, ,\cr
&j_n(\sigma,
\omega\beta^2,\beta^2)|_{\beta^2\to 0}\to {\pi^n\, \beta^{2 n}\over
(1+2\sigma)^{2 n}}\ {\Gamma(2\omega+n)\over n!\ \Gamma(2 \omega)}\ .}}
Also, assuming \main\ valid one has for the ratio \halpha\ 
\eqn\maincl{{\cal H}(\sigma/\beta+\omega\beta)
\big|_{\beta^2\to 0}\to -{m^2\over 16\, \sigma^2\, (1-4\sigma^2)^2}\ .}
Finally, $\mu|_{\beta^2\to 0}\to {m^2\over 16\pi \beta^2}$, and with 
\vevcl, \intcl, \maincl\ the expansion \reewwe\ takes the form
\eqn\limexp{\eqalign{&{{\langle\,
e^{i\omega\beta\varphi}(x)\,e^{i{\sigma\over\beta}
\varphi}(y)\,\rangle_{SG}}
\over{\langle\, e^{i\omega\beta\varphi}\,\rangle_{SG}\ \langle\, 
e^{i{\sigma\over\beta}\varphi}\,\rangle_{SG}}}\bigg|_{\beta^2\to 0}\to
\Big({t\over 8}\Big)^{4\omega\sigma}\ \Big[\gamma\big({1\over 2}-
\sigma\big)
\Big]^{2\omega}\ \bigg\{1-
{t^4\over 64}\ {\omega\,
(2\sigma+\omega)\over (1-4\sigma^2)^2}+\cr
&O(t^8)+
\ \ \sum_{n=1}^{\infty} {\Gamma(2\omega+n)\over n!\ \Gamma(2\omega)}\
\bigg[\, {2\,\gamma\big({1\over 2}-\sigma\big)\over
1+2\sigma}
\, \bigg]^{2 n}\Big({t\over 8}\Big)^{2 n(1+2\sigma)}\ \Big(1+
O\big(t^4\big)\, \Big)+\cr
&\ \ \sum_{n=1}^{\infty} {\Gamma(-2\omega+n)\over n!\ \Gamma(-2\omega)}\
\bigg[\, {2\, \gamma\big({1\over 2}+\sigma\big)\over
1-2\sigma}
\, \bigg]^{2 n}\Big({t\over 8}\Big)^{2 n(1-2\sigma)}\ \Big(1+
O\big(t^4\big)\, \Big)\, \bigg\}\ ,}}
where again $t=m\, |x-y|$. It is not difficult to see now that \penexp\ and
\limexp\ are in exact agreement with \semcorr. Thus the semiclassical
relation \semcorr\  actually yields \maincl\ and therefore supports our
main result \main. It is interesting to notice that agreement
between \limexp\ and \double\ further suggests the following explicit
expressions for some of the coefficients $B_{m,n}$ in \double,
\eqn\ksudyt{\eqalign{&
B_{0,n}={4^n\over n (1-2\sigma)^{2n}}\ \gamma^{2 n}\big(
{1\over 2}+\sigma\big)\, , \ \ \ \ \ n=1,2,\ldots\, ,\cr
&B_{m, 0}=B_{m+2,1}=0\, ,\ \ \ \ \ \ \ \ m=1,2,\ldots\ ,}}
which are not immediately obvious from the recursion relations of \ZAL.

\newsec{Comparison to perturbation theory}

\vskip 0.2in

It is easy to see that \main\ admits power series expansion in
$\beta^2$. For further references, let us quote two special cases. 
First, for $\alpha=0$
\eqn\exx{\langle\, (\partial\varphi)^2({\bar\partial}\varphi)^2\,
\rangle_{SG}=
-{{\pi^2 m^4}\over{4\sin^2 (\pi\xi)}} =
-{{m^4}\over
{4\beta^4}}+{{m^4}\over{2\beta^2}}+O(1)\ .}
Here we have chosen to express the VEV \maino\ through the mass
$m=2M\sin(\pi\xi/2)$ of the lightest sine-Gordon breather. Second, if
$\alpha=\omega\beta$ where $\omega$ is a constant,
\eqn\spt{\langle\, (\partial\varphi)^2({\bar\partial}\varphi)^2 \,
e^{i\beta\omega\varphi}\, \rangle_{SG} = -{{m^4}\over{4\beta^4}}\ 
{1\over
{1-4\omega^2}} + O\big({1\over {\beta^2}}\big)\ .}
These expansions can be compared with the results obtained directly
from \sg\ by means of ordinary Feynmann perturbation theory.

In developing the perturbation theory it is convenient to start from the
action \sg \ in its ``bare'' form
\eqn\sgbare{\eqalign{{\cal A}_{SG} =
&\int\ d^2 x\ {1\over{8\pi}}\biggl\{\ {1\over 2}
(\partial_\nu \varphi)^2
-{{m_{0}^2}\over{\beta^2}}\,\big[ \cos( \beta\varphi )\big]_B \
\biggr\}\  = \cr 
&\int\ d^2 x\ {1\over{8\pi}}\biggl\{\ {1\over 2}
(\partial_\nu \varphi)^2
+{{m_{0}^2}\over{2}}\,\big[\varphi^2\big]_B - {{m_{0}^2 \beta^2}\over {4!}}
\,\big[\varphi^4\big]_B + ... \
\biggr\}\ ,}}
where the symbol $[...]_B$ signifies that we are dealing with the bare
fields (as opposed to the renormalized fields defined in Sect.2), and it
is assumed that the action \sgbare \ is supplemented with
some cutoff procedure, with the cutoff momentum $\Lambda$. The scale
dimensions of the bare fields coincide with their naive values and hence
the bare mass parameter $m_0$ has the dimension of mass. The relation
between $m_0$ and the physical mass $m$ of the lightest breather
particle of \sg \ can be found perturbatively, order by order in
$\beta^2$. For instance, with the account of the leading mass correction
diagram in Fig.1, we have
\eqn\mmo{m_{0}^2 = m^2 + m^2 \beta^2 L  + O(\beta^4) \ ,}
where 
\eqn\logg{L={1\over \pi}\int{{d^2 k}\over {k^2 +
m^2}}=\log\big({{\Lambda^2}\over {m^2}}\big) +C \ ,}
and $C$ is a constant whose exact value depends on the implementation 
of the cutoff procedure. In fact, to all orders in $\beta^2$ this
relation has the form \Zarn \foot{Exact form of the function
$h(\beta^2)$ can be found in \Zarn.}
\eqn\mmoo{m_{0}^2 = m^2 \, e^{\beta^2 L}\ h(\beta^2); \qquad h(\beta^2) =
1+{{\pi^2}\over 6}\beta^4 +O(\beta^6) \ .}
Similarly, the relation between the bare exponential fields
$\big[e^{i\alpha\varphi}\big]_B$ and corresponding renormalized fields
can be written as 
\eqn\baref{ e^{i\alpha\varphi}= \bigg({1\over 4}\, \Lambda^2 \, e^{2\gamma
+C}\bigg)^{\alpha^2}\, \big[e^{i\alpha\varphi}\big]_B \ ,}
where $\gamma$ is Euler's constant and the normalization of the
field $e^{i\alpha\varphi}$ is fixed by the short-distance asymptotic
condition 
\eqn\fnorm{\langle\,  e^{i\alpha\varphi}(x)\
e^{-i\alpha\varphi}(y)\, \rangle_{SG} \to |x-y|^{-4\alpha^2} 
\qquad {\rm as} \qquad 
|x-y| \to 0 \ .}
 
Now we are prepared to make some perturbative calculations of the
one-point VEV. As the first example let us consider the field
$(\partial\varphi)^2 ({\bar\partial}\varphi)^2$. According to our
discussion in Sect.2 this renormalized field differs from the
corresponding bare field by appropriate counterterms
\eqn\renb{(\partial\varphi)^2 ({\bar\partial}\varphi)^2 = 
\big[(\partial\varphi)^2 ({\bar\partial}\varphi)^2\big]_B +
m_{0}^4\ \big(A_0 + A_1 \big[\cos(2\beta\varphi)\big]_B \big) +
m_{0}^2\ A_2 
\partial {\bar\partial}\big[\cos(\beta\varphi)\big]_B \ ,}
where $A_0 , A_1, A_2$ are constants which can depend on
$\beta$. In \renb \ the counterterms of the type 
$\Lambda^2 \big[\partial\varphi {\bar\partial}\varphi \big]_B$ and
$\Lambda^4$ which are needed to absorb the quadratic divergences in the
matrix elements of $\big[(\partial\varphi)^2
({\bar\partial}\varphi)^2\big]_B$ are not written down. In the following
calculations of these matrix elements we will systematically subtract
all quadratic divergences; with this convention the ``quadratic''
counterterms can be ignored altogether. The counterterms explicitly
shown in \renb \ are to compensate for remaining logarithmic
divergences. It is possible to see that this compensation can not be
achieved with the coefficient $A_2$ being just constant; 
instead one has to set
$A_2 = A\,L +B$, where $L$ is the logarithm \logg. The reason for this
subtlety lays in the fact that the field 
$(\partial\varphi)^2 ({\bar\partial}\varphi)^2$ always has a first
order resonance with the field $\partial{\bar
\partial}\cos(\beta\varphi)$ which results in the logarithmic scaling of
all its matrix elements which receive contributions from the above total
derivative field. Fortunately, here we are interested only in the
one-point VEV
\eqn\renv{\langle\, (\partial\varphi)^2
({\bar\partial}\varphi)^2\, \rangle_{SG} = 
\langle\, \big[(\partial\varphi)^2
({\bar\partial}\varphi)^2\big]_B\, \rangle_{SG} + A_0 m^4 e^{2\beta^2 L}\,
h^2 (\beta^2) + A_1 m^4 e^{-2\beta^2 L}\,  h^2 (\beta^2)\, g(2\beta) \ ,}
which gets no contribution from the last counterterm in \renb\ and
hence is not sensitive to the above subtlety. In writing \renv\  we have
used \mmoo\ to express $m_0$ through the physical mass and also used the
notation
\eqn\vex{\langle\, \big[e^{i\alpha\varphi}\big]_B\, \rangle_{SG} =
e^{-\alpha^2 L}\,g(\alpha)\ .}
Explicit expression for $g(\alpha)$ can be found in \LZ; here we will only
use the fact that for fixed $\omega$ 
\eqn\expg{g(\omega\beta)= 1+O(\beta^6)\ .}
The first term in \renv \ can be calculated directly using Feynmann
diagrams for \sgbare. To the leading order in $\beta^2$ one obtains
\eqn\lead{\langle\, \big[(\partial\varphi)^2
({\bar\partial}\varphi)^2\big]_B\, \rangle_{SG} = {1\over 2}\, m^4 L^2 +
O(\beta^2) \ .}
(Let us remind that we subtract the quadratic divergences). In order to
compensate for this $L^2$ divergence the counterterm coefficients in
\renv\ have to be chosen as follows
\eqn\as{A_0 = -{1\over {8\beta^4}}+O\big({1\over \beta^2}\big), \qquad 
A_1 = -{1\over {8\beta^4}}+O\big({1\over \beta^2}\big)}
and we obtain
\eqn\ford{\langle\, (\partial\varphi)^2
({\bar\partial}\varphi)^2\,\rangle_{SG} = -{{m^4}\over
{4\beta^4}}+O\big({1\over\beta^2}\big)\ . }
The calculation can be easily extended to the next order in $\beta^2$.
The next perturbative contribution to the VEV \lead\ comes from the
diagram in Fig.2. It has the form
\eqn\sublead{m^4 \beta^2\ (a_1 L^2 + a_2 L + a_3)\ ,}
where $a_1, a_2, a_3$ are numerical coefficients. It is not difficult to
check that in order to find the next term in \exx\ one only needs to
know the coefficient $a_1$ in front of the leading logarithmic term in
\sublead. This coefficient is evaluated directly from the diagram,
$a_1 = 1$. Then compensation of this term requires the following terms
in the $\beta^2$ expansion \as\ 
\eqn\ass{A_0 = -{1\over {8\beta^4}}+{1\over{4\beta^2}}+O(1)\, , \qquad 
A_1 = -{1\over {8\beta^4}}+{1\over{4\beta^2}}+O(1)\ .}
The finite terms remaining in \renv\ after the cancelation of the
divergences yield exactly \exx.

Next, let us apply the perturbation theory to more general VEV \vvv\
with $\alpha \neq 0$. Again, the renormalized field is a combination of
corresponding bare field and suitable counterterms,
\eqn\reb{\eqalign{ 
\big(\Lambda^2
e^{2\gamma+C}/4\big)^{-\alpha^2}\ (\partial\varphi)^2
&({\bar\partial}\varphi)^2\,e^{i\alpha\varphi}=
\big[(\partial\varphi)^2
({\bar\partial}\varphi)^2\,e^{i\alpha\varphi}\big]_B + \cr
&m_{0}^2\
\partial{\bar\partial}\big(B_{+}\,\big[e^{i(\alpha+\beta)\varphi}\big]_{B}
+ B_{-}\,\big[e^{i(\alpha-\beta)\varphi}\big]_{B}\big) + \cr
&m_{0}^4\ \big(A_{+}\,\big[e^{i(\alpha+2\beta)\varphi}\big]_{B} + 
A_{0}\, \big[e^{i\alpha\varphi}\big]_{B} +
A_{-}\,\big[e^{i(\alpha-2\beta)\varphi}\big]_{B} \big)\ .}}
The constants $A,B$ have to be determined from the requirement that the
renormalized field $(\partial\varphi)^2
({\bar\partial}\varphi)^2\,e^{i\alpha\varphi}$ has definite scale
dimension $4+2\alpha^2$. It is convenient to divide \reb\ by the VEV
\vex, and trade the parameter $m_0$ in favor of $m$,
\eqn\normb{\eqalign{&{{(\partial\varphi)^2
({\bar\partial}\varphi)^2\,e^{i\alpha\varphi}}\over{\langle\,
e^{i\alpha\varphi}\, \rangle_{SG}}} = {{\big[(\partial\varphi)^2
({\bar\partial}\varphi)^2\,e^{i\alpha\varphi}\big]_B}\over {\langle\,
\big[ e^{i\alpha\varphi}\big]_B\, \rangle_{SG}}} + \cr
&m^2\, h(\beta^2)\ \partial {\bar\partial} \bigg( B_{+}\, e^{-2\alpha\beta
L}\ {{g(\alpha+\beta)} \over {g(\alpha)}}\ \big[e^{i(\alpha+\beta)\varphi}
\big]_{N} + B_{-}\, e^{2\alpha\beta L}\ 
{{g(\alpha-\beta)} \over {g(\alpha)}}\ \big[e^{i(\alpha-\beta)\varphi}
\big]_{N}\bigg) + \cr
&m^4\, h^2 (\beta^2)\ \bigg( A_{+}\,e^{-(2\beta^2+4\alpha\beta)L}\  
{{g(\alpha+2\beta)}\over{g(\alpha)}}\
\big[e^{i(\alpha+2\beta)\varphi}\big]_{N} + 
A_{0}\, e^{2\beta^2 L} +\cr 
&A_{-}\,e^{-(2\beta^2-4\alpha\beta)L}\  
{{g(\alpha-2\beta)}\over{g(\alpha)}}\
\big[e^{i(\alpha-2\beta)\varphi}\big]_{N}\bigg)\ ,}}
where we used the notation
\eqn\normexp{\big[ e^{i\alpha\varphi} \big]_{N} =
{{e^{i\alpha\varphi}}/{\langle\, e^{i\alpha\varphi}\, \rangle_{SG}}} = 
{{\big[e^{i\alpha\varphi}\big]_{B}}/{\langle\,
\big[e^{i\alpha\varphi}\big]_{B}\,\rangle_{SG}}}\ .}
One can calculate perturbatively matrix elements of \normb\ , adjusting
the coefficients order by order in $\beta$ to ensure the cancelation of
all $L$-dependent terms. Note that \normexp\ contains no divergences and
so all the $L$ dependence of the counterterm part in \normb\ is shown 
explicitly. It turns out that contrary to the case $\alpha=0$ studying
just the VEV of \normb\ is not enough to determine the coefficients
$A_{+}, A_{0}, A_{-}$. We have considered the matrix elements of \normb\
between the vacuum and one- and two-particle states (involving the
lightest breather) along with the VEV. The calculations are straightforward
but rather bulky and we do not present them here. In the case $\alpha =
\omega\beta, \quad \omega\sim 1$ and in the leading order in $\beta^2$
the cancelation of $L$-dependent terms requires the following choice of
the coefficients,
$$
B_{+}={1\over{\beta^4}}\ {1\over{2\omega(1+\omega)^2}}+ 
O\big({1\over\beta^2}\big)\, ,\qquad
B_{-}=-{1\over{\beta^4}}\ {1\over{2\omega(1-\omega)^2}}+
O\big({1\over\beta^2}\big)\, ,
$$
and
\eqn\coffs{\eqalign{A_{+}= -{1\over{\beta^4}}\
{1\over {16(1+\omega)(1+2\omega)}}+
&O\big({1\over\beta^2}\big)\, , \qquad 
A_{-}=  -{1\over{\beta^4}}\ {1\over {16(1-\omega)(1-2\omega)}}
+O\big({1\over\beta^2}\big)\, , \cr
&A_{0}=-{1\over{\beta^4}}\ {1\over{8(1+\omega)(1-\omega)}}+
O\big({1\over\beta^2}\big)\ .}}
With \coffs\ the result for the VEV of this field identical to \spt\ 
immediately follows from \normb.

\newsec{Exact results from XYZ model}

\vskip 0.2in

As is well known \luth, the sine-Gordon QFT \sg\ can be obtained by taking
an appropriate scaling limit of the XYZ spin chain described by the 
Hamiltonian
\eqn\xyz{
{\bf H}_{XYZ}=-{1\over{2\varepsilon}}\sum_{s=1}^{N}
\big(J_{x}\,\sigma_{s}^{x}
\, \sigma_{s+1}^{x}+
J_{y}\,\sigma_{s}^{y}
\, \sigma_{s+1}^{y}+J_{z}\,\sigma_{s}^{z}
\, \sigma_{s+1}^{z}-J\big) \ ,}
with $J_{x}\geq J_{y}\geq |J_{z}|$. In \xyz\ we have 
introduced an auxiliary
parameter $\varepsilon$ which is interpreted as a lattice spacing. It is
convenient to use the Baxter's elliptic parameterization \Baxter\ of the
coefficients $J$ in \xyz,
\eqn\psosi{\eqalign{&J_x={1-\beta^2\over \pi }\biggl(
{\theta_{4}(\beta^2) \theta'_{1}(0)\over
\theta_{4}(0) \theta_{1}(\beta^2)}+
{\theta_{1}(\beta^2) \theta'_{1}(0)\over
\theta_{4}(0) \theta_{4}(\beta^2)} \biggr)\ ,\cr
&J_y={1-\beta^2\over \pi }\biggl(
{\theta_{4}(\beta^2) \theta'_{1}(0)\over
\theta_{4}(0) \theta_{1}(\beta^2)}-
{\theta_{1}(0) \theta'_{1}(0)\over
\theta_{4}(0) \theta_{4}(\beta^2)} \biggr)\ ,\cr
&J_z={1-\beta^2\over \pi }\biggl(
{\theta'_{1}(\beta^2)\over \theta_{1}(\beta^2)}-
{\theta'_{4}(\beta^2)\over \theta_{4}(\beta^2)}  \biggr)\ ,\cr
&J=-{1-\beta^2\over \pi }\biggl(
{\theta'_{1}(\beta^2)\over \theta_{1}(\beta^2)}+
{\theta'_{4}(\beta^2)\over \theta_{4}(\beta^2)}  \biggr)\ , }}
where
$$\eqalign{&\theta_{1}(v)=2 p^{{1\over 4}}\   {\rm sin}(\pi v)\
\prod_{n=1}^{ \infty} \big(1-p^{2 n}\big)\, \big(1-e^{2\pi i\, v}\,
p^{2 n}\big)\,
\big(1-e^{-2\pi i\, v}\,p^{2 n}\big)\ ,\cr
&\theta_{4}(v)=
\prod_{n=1}^{ \infty} \big(1-p^{2 n}\big)\, \big(1-e^{2\pi i\, v}\,
p^{2 n-1}\big)\,
\big(1-e^{-2\pi i\, v}\,p^{2 n-1}\big)\ }$$
and the prime in\ \psosi\ denotes a derivative. The scaling limit of
\xyz\ is achieved by sending
\eqn\sclim{N\to\infty\, ,\ \ \ \ \ \ \ \varepsilon\to 0\, ,\ \ \ \ \ \ \
p\to 0\,}
with the combinations
\eqn\scpar{R=N\, \varepsilon  \, ,  \ \ \ \ 
M = {4\over\varepsilon}\ p^{(1+\xi)/4}}
kept fixed. According to Refs.\refs{\Jons, \luth} in this
limit the energy spectrum of \xyz\ is described by the QFT \sg, the 
parameter $M$ coinciding with the sine-Gordon soliton mass.

In fact, the QFT \sg\ itself controls only the leading $p\to 0$
singularities in the spectrum of \xyz. Using exact XYZ ground state
energy \Baxter\ one can easily extract subleading singular terms in this
quantity. Being expressed through the scaling parameters $M$ and $R$, the
singular at $p\to 0$ part of the bulk ground state energy reads
\eqn\jshdyt{  (E_{XYZ})_{sing}=-{R M^2\over 4}\ \tan(\pi
\xi/2)
\ \bigg\{\, 1+ \Big({M\varepsilon\over 4}\Big)^2+
O(\varepsilon^4)\, \bigg\}\, \ \ \ \ \  (\, RM\gg 1\, )\ .}
Whereas the leading term here is exact sine-Gordon vacuum energy, the
higher-order in $\varepsilon$ terms must be attributed to the irrelevant
operators which differ the exact XYZ Hamiltonian \xyz\ from the
Hamiltonian ${\bf H}_{SG}$ of the sine-Gordon QFT \sg. As follows from
the analysis in \SL, for $\beta^2 < 2/3$ the leading in $\varepsilon$
correction comes from the terms
\eqn\hsgdt{{\bf H}_{XYZ}=const+ {\bf H}_{SG}-{{\varepsilon^2}\over 16}\
\int_{0}^R {d x\over 2 \pi}\,
\Big( \lambda_{+}\, (\partial\varphi)^2 ({\bar\partial}\varphi)^2 + 
\lambda_{-}\, \big( (\partial\varphi)^4 +({\bar\partial}\varphi)^4 
\big )\Big)+\ldots\  ,}
where $\lambda_{+}$ and $\lambda_{-}$ are numerical coefficients whose
exact values are found in \SL\  and 
the dots stand for the irrelevant operators of higher
dimensions. The corrections in \jshdyt\ can be expressed through the 
expectation
values of the correction terms in \hsgdt\ over the sine-Gordon vacuum. 
Obviously, it is the VEV of the operator 
$(\partial\varphi)^2 ({\bar\partial}\varphi)^2$ which is responsible for
the $\varepsilon^2$ term in \jshdyt\  (the VEV of $(\partial\varphi)^4$
and $({\bar\partial}\varphi)^4$ vanish), i.e.
\eqn\encor{\lambda_{+}\ \langle\, (\partial\varphi)^2
({\bar\partial}\varphi)^2\, \rangle_{SG} = {{M^4}\over 4}\tan(\pi\xi/2)\
.}
Using the result of \SL\ 
\eqn\jsdhgyt{
\lambda_+=-{\cot(\pi\xi/2)\over 2 \pi}\ ,}
one arrives precisely at the Eq.\maino\ which therefore agrees with exact
results of the lattice theory.

\newsec{Application: Two-point correlation function in scaling Lie-Yang model}

\vskip 0.2in

In Ref.\ZamAl\ a two-point correlation function in so called Scaling
Lee-Yang Model (SLYM) was studied. In particular, a combination of the
operator product expansions and conformal perturbation theory was used
there to develop a short-distance expansion similar to \reewwe. 
In this section we will use our result \evllk\ to extend this expansion
further thus obtaining more accurate estimate for the two-point
correlation function at all distances.

The SLYM is one of the simplest of the perturbed CFT \mpaa, namely
\eqn\ksidy{{\cal A}_{SLYM}=
{\cal M}_{2/5}+i h\, \int d^2 x \,  \Phi(x)\ ,}
where
$$\Phi(x)=\Phi_{1,3}(x)\, ,\ \ \ \ \Delta_{\Phi}=-{1\over 5}\ .$$
As is known (see e.g. \Muss) the QFT \ksidy\ is massive; it has one sort
of massive particles whose mass $m$ is related to the parameter $h$ in 
\ksidy\ as  
\eqn\bdgfr{h={2^{{1\over 5}}\,
5^{{3\over 4}}\over 16\, \pi^{{6\over 5}} }\
{\big(\Gamma(2/3)\, \Gamma(5/6)\big)^{{12\over 5}}\over
\Gamma(3/5)\, \Gamma(4/5)}\ m^{{12\over 5}}=0.0970485\ldots\,
m^{{12\over 5}}\ .}
We will use the notations
\eqn\hsgdr{
\Theta(x)=T_{\nu}^{\nu}(x)/4=i h\pi\,  (1-\Delta_{\Phi})\ \Phi(x)\ }
for the trace of the energy-momentum tensor associated with \ksidy. 

Consider the two-point correlation function 
\eqn\hsydtr{G(r)=\langle\,  \Theta(x) \,
\Theta(0)\,  \rangle\, , \ \ \ \ \ \ \  r=|x|\ . }
According to \ZamAl\ this correlation function admits the following
short-distance expansion
\eqn\hsgdytt{\eqalign{G(r)=&-h^2\pi^2\,
(1-\Delta_{\Phi})^2\ C_{\Phi\Phi}^{I}(r)+
i h\pi\,  (1-\Delta_{\Phi})\   C_{\Phi\Phi}^{\Phi}(r)\
\langle\,\Theta\,\rangle- \cr
&h^2\pi^2\,  (1-\Delta_{\Phi})^2\ C_{\Phi\Phi}^{{T{\bar T}}}(r)\
\langle\,T\bar{T}\,\rangle+O\big(r^{{42\over 5}}\big)\  ,}}
where the notation
\eqn\ttbard{T{\bar T} = L_{-2}{\bar L}_{-2}I}
is used. The coefficient functions $C$ in \hsgdytt\ admit power series
expansions in $h$, the first few terms being known explicitly \ZamAl\
\foot{Notice the analytic expression for the first subleading term in
the expansion of $C_{\Phi\Phi}^{\Phi}$, which was given numerically in
\ZamAl.}
\eqn\hsyduur{\eqalign{&C_{\Phi\Phi}^{I}(r)=r^{{4\over 5}} \, \Big\{1+
{ 5^{{1\over 4}}\over 1960}\, {\Gamma^4(1/5) \, \Gamma(3/5)\over
\Gamma^3(4/5)}\ hr^{12\over 5}+O\big(r^{{24\over 5}}\big)\ \Big\}\, ,\cr
&C_{\Phi\Phi}^{\Phi}(r)=i\ {5^{{1\over 4}}\over 10\pi}\
{\Gamma^2(1/5) \, \Gamma(2/5)\over
\Gamma(4/5)}\ r^{{2\over 5}} \
\Big\{1+ {2\, \pi^2\over  5^{{13\over 4}}\, 9 }\
{\Gamma^2(1/5)\, \Gamma^2(2/5)\over
\Gamma^3(3/5)\, \Gamma^3(4/5)}\ hr^{12\over 5}+O\big(r^{{24\over 5}}\big)\
\Big\}\, ,
\cr
&C_{\Phi\Phi}^{T{\bar T}}(r)={r^{{24\over 5}}\over 121}\
\Big\{1+
O\big(r^{{12\over 5}}\big)\ \Big\}\, .}}
With the known exact VEV of the field $\Theta$,
\eqn\tvev{\langle\,\Theta\,\rangle=-{\pi \over 4\sqrt3}\ m^2\, ,}
Eqs.\hsgdytt, \hsyduur\  effectively gives the short-distance
expansion of the correlation function \hsydtr\ up to the terms 
$\sim r^{16\over 5}$ \ZamAl. Now, using \evllk\ we can derive the VEV
\eqn\ttbare{\langle\,T\bar{T}\,\rangle=-{\pi^2 \over 48}\ m^4\ .}
This additional peace of data allows one to compute explicitly the term
$\sim r^{24\over 5}$ in \hsgdytt. The next term $\sim r^{26\over 5}$,
which would come from the $h^2$ term in $C_{\Phi\Phi}^{\Phi}$, is still
not available in an analytic form.

The correlation function \hsydtr\ admits also the large-distance
expansion in terms of exact form factors \ZamAl. Two leading terms,
corresponding to zero- and one-particle contributions, are known in
analytic form,
\eqn\jshdtr{G(r)= {\pi^2\over 48}\, m^4\,  \Big\{1-{27\over 10\pi^2}\ Z
\, K_0(mr)+\ldots\ \Big\}\ ,}
where
$$Z={10 \sqrt3 \pi\over 27}\ \exp\Big\{-\int_0^{{2\pi\over 3}}\,
{dt\over \pi}\ {t\over \sin(t)}\ \Big\}=0.8155740\ldots\ , $$
and $K_0(t)$ is the MacDonald function. Further terms in this expansion
of $G(r)$ can be obtained by numerical integration of its spectral
representation including the contributions of two or more particles in
the intermediate state \ZamAl. The expansion is known to converge very
fast. With the inclusion of up to four-particle contributions this
expansion gives a precision better then $10^{-2}\%$ for $mr \geq 10^{-2}$.
The short-distance expansion \hsgdytt\ (with \hsyduur, \tvev\ and
\ttbare) is compared with this data in Table 1. The combined data
from these two expansions apparently have relative precision $10^{-5}\%$
or better for all values of $r$. 

Finally let us note that since exact form factors of the sine-Gordon model
are known \ \refs{\Fedya,\lik}, similar numerical analysis can be 
performed for the general sine-Gordon correlation function \twopoint.

\hskip2.0cm

\centerline{\bf Acknowledgments}

\hskip0.5cm

A.Z. acknowledge kind hospitality of ITP at Santa Barbara (and the
supporting NSF grant No. PHY94-07194) where this
work was started. Research work of S.L. and A.Z. is 
is supported by DOE grant \#DE-FG05-90ER40559.
\vfill
\eject

\midinsert
{
\baselineskip=10pt
\centerline{
\noindent\vbox{\offinterlineskip
\def\tablerule{\noalign{\hrule}}
\halign{
\vrule height8.5pt depth3.5pt width0pt #&\vrule#\tabskip=1em plus2em&
   #&\vrule#&
   #&\vrule#&
   #&\vrule#&
   #&\vrule#
\tabskip=2pt
\cr\tablerule
&&
&& Long-distance expansion
&& &\omit&  \hidewidth Short-distance expansion  &
\cr\tablerule
&& $mr$  &&\  0-1-2-3-4 particles  && Without $\langle\, T\bar{T}\,\rangle$
 && With $\langle\,T\bar{T}\,\rangle$&
\cr\tablerule
&& 0.001  &&\   0.01964182  &&\ 0.01947405  && 0.01947405 &
\cr\tablerule
&& 0.002  &&\   0.02553233  &&\ 0.02547150  && 0.02547150 &
\cr\tablerule
&& 0.005  &&\   0.03616744  &&\ 0.03615515  && 0.03615515 &
\cr\tablerule
&& 0.010  &&\   0.04689571  &&\ 0.04689286  && 0.04689286 &
\cr\tablerule
&& 0.020  &&\   0.06045855  &&\ 0.06045806  && 0.06045806 &
\cr\tablerule
&& 0.040  &&\   0.07730747  &&\ 0.07730741  && 0.07730741 &
\cr\tablerule
&& 0.060  &&\   0.08881026  &&\ 0.08881025  && 0.08881025 &
\cr\tablerule
&& 0.080  &&\   0.09771443  &&\ 0.09771443  && 0.09771443 &
\cr\tablerule
&& 0.100  &&\   0.10502922  &&\ 0.10502922  && 0.10502922 &
\cr\tablerule
&& 0.120  &&\   0.11125352  &&\ 0.11125352  && 0.11125352 &
\cr\tablerule
&& 0.140  &&\   0.11667524  &&\ 0.11667523  && 0.11667525 &
\cr\tablerule
&& 0.160  &&\   0.12147752  &&\ 0.12147749  && 0.12147752 &
\cr\tablerule
&& 0.180  &&\   0.12578494  &&\ 0.12578489  && 0.12578495 &
\cr\tablerule
&& 0.200  &&\   0.12968660  &&\ 0.12968651  && 0.12968661 &
\cr\tablerule
&& 0.220  &&\   0.13324863  &&\ 0.13324850  && 0.13324866 &
\cr\tablerule
&& 0.240  &&\   0.13652170  &&\ 0.13652150  && 0.13652174 &
\cr\tablerule
&& 0.260  &&\   0.13954552  &&\ 0.13954523  && 0.13954559 &
\cr\tablerule
&& 0.280  &&\   0.14235191  &&\ 0.14235151  && 0.14235201 &
\cr\tablerule
&& 0.300  &&\   0.14496678  &&\ 0.14496622  && 0.14496693 &
\cr\tablerule
&& 0.400  &&\   0.15581014  &&\ 0.15580799  && 0.15581079 &
\cr\tablerule
&& 0.500  &&\   0.16401350  &&\ 0.16400745  && 0.16401562 &
\cr\tablerule
&& 0.600  &&\   0.17045818  &&\ 0.17044413  && 0.17046372 &
\cr\tablerule
&& 0.700  &&\   0.17564907  &&\ 0.17562051  && 0.17566157 &
\cr\tablerule
&& 0.800  &&\   0.17990480  &&\ 0.17985216  && 0.17993010 &
\cr\tablerule
&& 0.900  &&\   0.18344031  &&\ 0.18335022  && 0.18348740 &
\cr\tablerule
&& 1.000  &&\   0.18640771  &&\ 0.18626233  && 0.18648980 &
\cr\tablerule
&& 1.200  &&\   0.19105758  &&\ 0.19072646  && 0.19127219 &
\cr\tablerule
&& 1.400  &&\   0.19446608  &&\ 0.19380550  && 0.19494924 &
\cr\tablerule
&& 1.600  &&\   0.19700864  &&\ 0.19581239  && 0.19798354 &
\cr\tablerule
&& 1.800  &&\   0.19892990  &&\ 0.19691763  && 0.20073902 &
\cr\tablerule
&& 2.000  &&\   0.20039614  &&\ 0.19720202  && 0.20353862 &
\cr\tablerule
&& 2.500  &&\   0.20275757  &&\ 0.19434989  && 0.21284363 &
\cr\tablerule
&& 3.000  &&\   0.20402333  &&\ 0.18570096  && 0.23007149 &
\cr\tablerule
\noalign{\smallskip} }
}
}
}
\capt{Table 1. Comparison of short and long-distance
expansions for the two-point correlation function \hsydtr\ . The first
column gives the results of long-distance expansion which includes
contributions of up to four-particle states (the four-particle
contribution which we include here represents the improvement over the
data in \ZamAl\ ). The data in the second and the third
columns correspond to the short distance expansion \hsgdytt\ without the
$T{\bar T}$ term and with this term, respectively.}
\endinsert
{\bf \centerline{Figures}}

\vskip 0.2in
\centerline{\epsfxsize 3.1truein\epsfbox{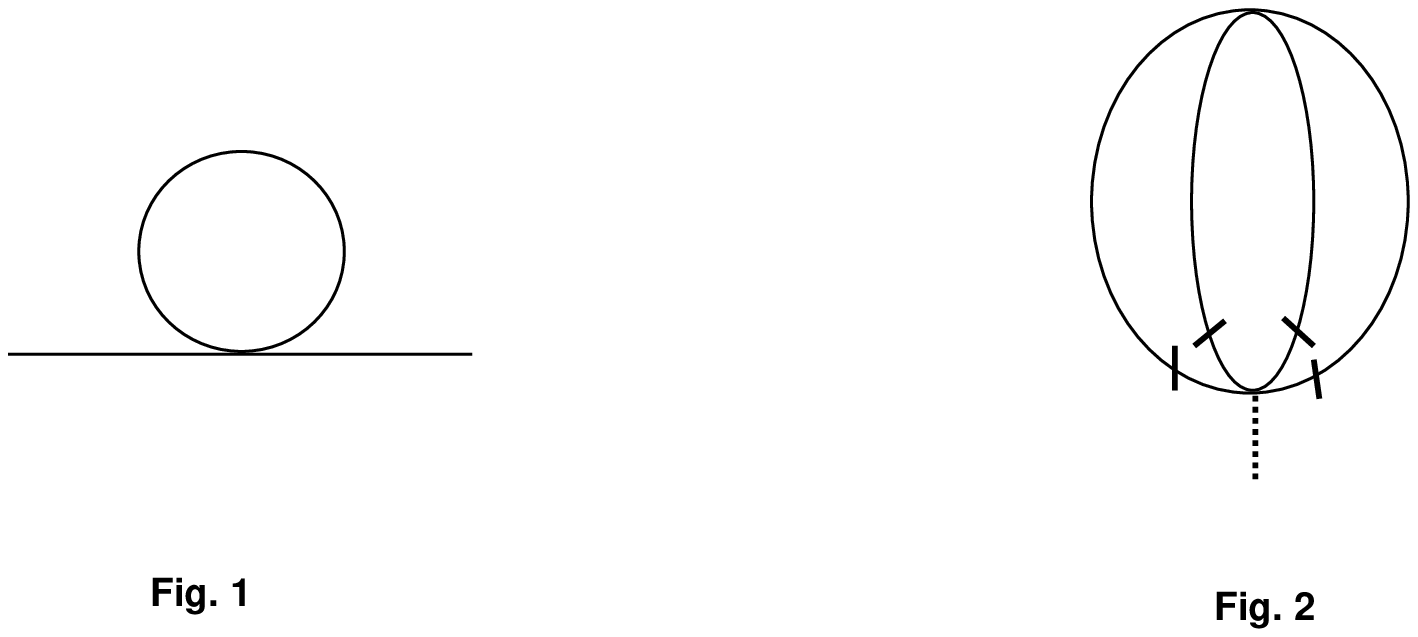}}
{\bf Fig.1.} The leading mass correction diagram which gives \mmo\ .

{\bf Fig.2.} Diagram contributing to the VEV \lead\ in the order
$\beta^2$. The strokes over the propagators stand for the
derivatives.

\vfill
\listrefs

\end